\documentclass[twocolumn]{aastex631}
\usepackage{amsmath}
\usepackage{graphicx}
\graphicspath{{./}{figures/}}
\usepackage{hyperref}

\usepackage[normalem]{ulem}
\usepackage{soulutf8}

\renewcommand{\Pr}{\mathrm{Pr}}
\newcommand{\zfront}{z_\mathrm{front}}
\newcommand{\approptoinn}[2]{\mathrel{\vcenter{
  \offinterlineskip\halign{\hfil$##$\cr
    #1\propto\cr\noalign{\kern2pt}#1\sim\cr\noalign{\kern-2pt}}}}}

\begin{document}

\title{3D Simulations Demonstrate Propagating Thermohaline Convection for Polluted White Dwarfs}

\author[0000-0002-4538-7320]{Imogen G.~Cresswell}
\affiliation{Department of Astrophysical \& Planetary Science, University of Colorado, Boulder, CO 80309, USA}

\author[0000-0003-4323-2082]{Adrian E.~Fraser}
\affiliation{Department of Astrophysical \& Planetary Science, University of Colorado, Boulder, CO 80309, USA}
\affiliation{Department of Applied Mathematics, University of Colorado, Boulder, CO 80309, USA}

\author[0000-0002-4791-6724]{Evan B.~Bauer}
\affiliation{Center for Astrophysics $|$ Harvard \& Smithsonian, 60 Garden Street, Cambridge, MA 02138, USA}

\author[0000-0002-3433-4733]{Evan H.~Anders}
\affiliation{Kavli Institute for Theoretical Physics, University of California, Santa Barbara, Santa Barbara, CA 93106, USA}

\author[0000-0001-8935-219X]{Benjamin P.~Brown}
\affiliation{Department of Astrophysical \& Planetary Science, University of Colorado, Boulder, CO 80309, USA}

\begin{abstract}
Polluted white dwarfs (WDs) with small surface convection zones deposit significant concentrations of heavy elements to the underlying radiative interior, presumably driving thermohaline convection. 
% Current models of polluted white dwarfs often do not account for this effect, which can increase the inferred accretion rate by orders of magnitude when it is included. 
Current models of polluted WDs frequently fail to account for this effect, although its inclusion can increase the inferred accretion rate by orders of magnitude.
It has been argued that this instability cannot be treated as a continuous mixing process and thus should not be considered in these models. In this work, we study three-dimensional simulations of a thermohaline-unstable layer propagating into an underlying stable region, approximating the polluted WD scenario. We find that although thermohaline convection works to reduce driving gradients somewhat, %stabilize the system, 
the front continues to propagate and the system remains unstable. Importantly, the turbulent flux of metals broadly dominates over the diffusive flux in quantitative agreement with with existing mixing prescriptions implemented in some stellar evolution models (except slightly below the boundary of the propagating front, where recent prescriptions neglect overshoot-like effects). Thus, our results broadly support polluted WD models that include thermohaline mixing in their estimates of the settling rate.
%indicating that the front continues to be pushed by advective flux of composition and is not simply diffusing away. We find bulk values in the 3D simulations are well predicted by the Brown model, but that it fails to capture subtlties at the boundary. \todo{just a rough draft}

\end{abstract}
%\maketitle

\section{Introduction}
Many white dwarfs show evidence of accretion from rocky exoplanetary material \citep{Zuckerman2003, Zuckerman2010, Koester2014, OuldRouis2024}. Remarkably, because white dwarfs otherwise have pristine atmospheres composed of only the lightest elements (hydrogen and/or helium), the spectral features from this accretion enable detailed measurements of the bulk composition of exoplanetary bodies. The strong gravity at the surfaces of WDs makes gravitational sedimentation efficient, so that heavy elements accreted from rocky material can sink away from the observable photosphere region on timescales ranging from days to millions of years, depending on the amount of convective mixing near the surface, which is a function of temperature and atmospheric composition \citep{Koester2009}. Less surface convection means a faster timescale for heavy elements to disappear from the surface. Observing heavy elements at the surfaces of these stars therefore implies very recent or ongoing accretion.

White dwarfs of type DA or DAZ have atmospheres dominated by hydrogen, and at effective temperatures above around 15{,}000\,K are expected to have negligible surface convection, leading to very short sedimentation timescales on the order of days. Models of WD surface layers that account for accretion and sedimentation of heavy elements mediated only by atomic plasma diffusion imply that this accretion would quickly lead to significant concentrations of heavy elements built up near the surface, which would lead to a composition gradient that should drive thermohaline convection \citep{Deal_WD,Wachlin2017}. One-dimensional (1D) stellar evolution models that account for thermohaline convection (or, ``thermohaline mixing") rely on approximations for a local, turbulent diffusivity based on the composition gradient such as those reported in \cite{kippenhahn_thermohaline} or \cite{Brown_2013}. WD models using these prescriptions and accounting for thermohaline convection in DAZ WDs with $T_{\rm eff} \gtrsim 15,000$\,K have inferred accretion rates several orders of magnitude higher than previously inferred without accounting for thermohaline convection \citep{bauer_increases_2018,bauer_polluted_2019,Wachlin_WDs,Dwomoh2023}.

\cite{Koester2015} raised objections to models that portray thermohaline convection as the dominant mixing process at the surfaces of these warm DAZ WDs, and much recent observational literature on these polluted WDs continues to either ignore thermohaline mixing or describe its significance as ``debated'' based on these arguments (e.g., \citealt{Rogers2024,Bedard2024,Williams2024}). Models of polluted WDs yield disagreements on inferred bulk composition and overall rate of mass accretion, the latter by orders of magnitude, motivating further calculations to provide a clear resolution to this debate. One of the primary objections that \cite{Koester2015} raised to the physical picture of thermohaline mixing as applied to polluted WD models is that these models treat it as a continuous mixing process, while the physical description presented in \cite{kippenhahn_thermohaline} is more akin to a one-time instability that acts to flatten the driving composition gradient over a local thermal timescale. \cite{Koester2015} objects that it is inappropriate to translate this into a diffusive mixing prescription that operates continuously over timescales longer than a discrete instability event driven by local thermal diffusion.

In this work, we directly address \cite{Koester2015}'s objection with 3D fluid simulations that incorporate a top boundary condition corresponding to continuous flux of heavy composition driving thermohaline instability and associated mixing over timescales much longer than than the local thermal diffusion timescale. This simulation setup closely mirrors the configuration of polluted white dwarfs that experience continuous accretion of heavy exoplanetary debris driving unstable composition gradients at their surfaces. These 3D simulations do not rely on any 1D effective diffusivity prescription such as that of \cite{kippenhahn_thermohaline}, but we can compare our simulations against 1D prescriptions and measure fluid velocities and composition fluxes. We show that with a continuous, externally supplied flux of heavy material, thermohaline instability can in fact drive continuous mixing for timescales much longer than local thermal diffusion timescales, forming many successive generations of the characteristic fingers that mediate the fluid motions and associated flux. It is not accurate to describe thermohaline instability as only capable of producing a ``one-time event", and our simulations validate the importance of incorporating ongoing thermohaline mixing into models of white dwarfs with composition gradients that are thermohaline-unstable. Our models therefore lend support to the higher accretion rates inferred for warm polluted DAZ WDs when accounting for thermohaline instability.

% \begin{itemize}
%     \item Polluted white dwarfs have heavier elements at the surface that we observe -- thermohaline mixing was not believed to be a process involved but this work may show that we should be considering it?
%     \item Polluted white dwarfs are white dwarfs that have engulfed a planet or asteroid and therefore have observable elements like magnesium, iron, and calcium when they are usually mainly carbon and oxygen (CO white dwarf) or oxygen, neon, and magnesium
%     \item these heavy elements are observable and therefore are not mixing instantly 
%     \item If we can show that the thermohaline front propagates and then stops at certain parameters but still has thermohaline unstable layer, this could support the theory that thermohaline mixing is occuring in these locations - a somewhat direct rebuttal to what the community thinks currently
%     \item Papers from Evan B:
% \end{itemize}

% \subsection{Thermohaline mixing is ...}

% \subsection{In this letter, we ...}
% \newpage

\begin{figure*}[t!]
     \centering
     \includegraphics[width=0.95\linewidth]{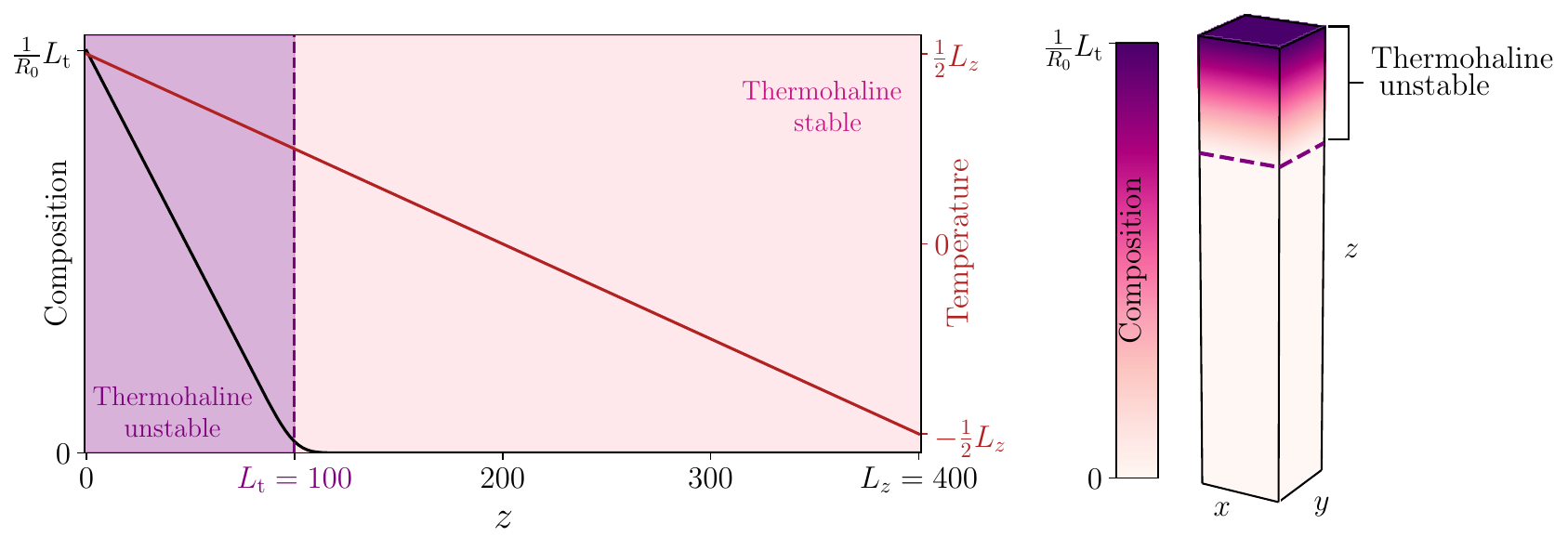}
     \caption{Initial composition (black line) and temperature (red line) profiles  (left) and 3D visualisation of initial composition field (right). The purple dashed line shows the initial location of the front at $z=100$, above which the system is thermohaline-unstable. 
     %Stable region $r_f > 1$ and $R_0 > 1/\tau$.
     }
     \label{params}
\end{figure*}

\section{Methods} \label{sec:methods}
Here we summarize our simulation configuration, choice of units, and numerical methods, with each subject discussed at greater length in Appendix \ref{sec:non-dimensionalization}.

We study the evolution of a stably-stratified (i.e., Ledoux-stable), Cartesian fluid layer with fluctuations in temperature $T$ and chemical composition $C$ (analogous to mean molecular weight $\mu$), and ignore the effects of rotation and magnetic fields. Our coordinate system is oriented such that $z$ is aligned with gravity and increases with depth. The layer is confined vertically between two plates, with a constant flux of high-$C$ material injected from above to approximate active accretion, a fixed temperature gradient at the top, and fixed $T$ and $C$ at the bottom of the domain (though we restrict our reported results to times before the bottom boundary condition has noticeable effects on the dynamics). 

Our initial $T$ and $C$ profiles are shown in Fig.~\ref{params}: our simulations begin with an initially uniform, stable (i.e., sub-adiabatic) temperature gradient throughout our domain, and a composition profile that varies linearly with depth in the top section of the domain (with slope matching the imposed $C$ flux at the boundary) and is uniform in the rest of the domain, with a smoothing function to transition between the two regions. Additionally, a low-amplitude white-noise perturbation is added to $T$ to seed instability at the start of our simulations. We assume that our domain height is smaller than a pressure scale height and that our flows are much slower than the local sound speed so that dynamics are well-captured by the Boussinesq approximation (\citealt{spiegel_boussinesq_1960}; note that our assumed small domain height also justifies the use of a Cartesian box). %\citep[see][and note that our assumed small domain height also justifies the use of a Cartesian box]{spiegel_boussinesq_1960}. 
% In this limit, the governing equations are
% \begin{gather}
% \nabla \cdotp \mathbf{u} = 0,\label{eq:divu}\\
% \frac{\partial \mathbf{u}}{\partial t} + \mathbf{u} \cdot \nabla \mathbf{u} = - \nabla P + \mathrm{Pr}(T - C)\mathbf{\hat{z}} + \mathrm{Pr}\nabla^{2}\mathbf{u}, \label{eq:mom} \\
% \frac{\partial C}{\partial t} + (\mathbf{u}\cdotp\mathbf{\nabla})C = \tau \nabla^{2}C , \label{eq:comp} \\
% \frac{\partial T}{\partial t} + (\mathbf{u}\cdotp\mathbf{\nabla})T = \nabla^{2}T , \label{eq:temp} \\
% \end{gather}
% where $\mathbf{u}$ is the fluid velocity, $P$ is the pressure perturbation, and $T$ and $C$ are the temperature and composition fluctuations about the background mean. 
% Here, we have nondimensionalized our system by measuring lengths in units of the characteristic width of fingers $d = (\kappa_T \nu/N_T^2)^{1/4}$ (with $\kappa_T$ the thermal diffusivity, $\nu$ the kinematic viscosity, and $N_T$ the local Brunt–V\"{a}is\"{a}l\"{a} frequency defined from the temperature stratification; \citealt{Stern_1960}), time with the corresponding thermal diffusion time, and temperature and composition based on the background thermal stratification. Further details are given in Appendix \ref{sec:non-dimensionalization}.
We simulate the resulting equations using the Dedalus code \citep{Dedalus_methods}. 

The fingers associated with thermohaline convection have a characteristic width that can be estimated in terms of basic fluid parameters \citep{Stern_1960,garaud_DDC_review}, and coupled with the thermal diffusivity this provides convenient units for both length and time scales for describing our simulations.
%Following, e.g, \citet{garaud_DDC_review}, we 
We therefore 
express lengths in units of the typical finger width $d$ which can be written in terms of the kinematic viscosity $\nu$, thermal diffusivity $\kappa_T$, and the local Brunt–V\"{a}is\"{a}l\"{a} frequency $N_T^2$ based on our initial temperature profile as
\begin{equation}
    [x] = d \equiv \left( \frac{\kappa_T \nu}{N_T^2} \right)^{1/4},
\end{equation}
and time in units of the associated thermal diffusion time,
\begin{equation}
    [t] = \frac{d^2}{\kappa_T}.
\end{equation}
In these units, our horizontal domain size is $L_x = L_y = 75$, our initial unstable composition gradient (ignoring the smoothed transition region) extends to a depth of $L_t = 100$, and the full domain height for most of our simulations is $L_z = 400$. To allow dynamics to evolve further before the artificial bottom boundary significantly impacts them, some runs use taller domains with $L_z = 500$. 
Note that, for all results presented here, we have shifted the definition of $t=0$ so that it represents the first timestep where the turbulent fluxes have reached a sufficient amplitude to adjust the initial composition profile. We express temperature in units such that the initial (potential) temperature profile has a gradient of $-1$, and composition in units such that a gradient of $-1$ is equivalent to marginal stability by the Ledoux criterion for convection. 
Full details of our equations, choice of units, and numerical methods are given in Appendix \ref{sec:non-dimensionalization}. %\textcolor{red}{[Put equations for $d$ and $[t]$ in here; also, put the stellar astrophysicist's definition of $R_0$ in, and the version in terms of Brunt frequencies]}

Two key dimensionless control parameters that we vary between our simulations are the Prandtl number $\Pr$ and diffusivity ratio $\tau$, defined by 
\begin{equation}
     \Pr \equiv \frac{\nu}{\kappa_T} \quad \mathrm{and} \quad \tau \equiv \frac{\kappa_C}{\kappa_T},
\end{equation}
%where $\nu$ is the kinematic viscosity, $\kappa_T$ the thermal diffusivity, and 
where $\kappa_C$ the compositional diffusivity. The values of both $\Pr$ and $\tau$ are typically much less than one in stars \citep[see, e.g.,][]{Garaud2015,bauer_increases_2018}, and are generally much smaller than what can be achieved with modern computing capabilities. The third key control parameter in our system is the imposed composition flux at $z=0$, and corresponding slope of the initial composition profile at the top of the domain. In our choice of units, this slope becomes $R_0^{-1}$, where
\begin{equation}
    R_0 \propto \frac{\partial_z T_0 - \partial_z T_\mathrm{ad}}{\partial_z C_0} \sim \frac{[\text{stabilizing $T$ gradient}]}{[\text{destabilizing $C$ gradient}]}
\end{equation}
(full definition in Eq.~[\ref{eq:control-parameters}]) is the \emph{density ratio}, a dimensionless number that measures the strength of the stabilizing temperature gradient relative to the destabilizing composition gradient. In terms of local properties typically calculated in stellar models, this quantity can be written \citep{Pascale_notes}
\begin{equation}
    R_0 = \frac{\nabla - \nabla_\mathrm{ad}}{\frac{\phi}{\delta} \nabla_\mu} = \frac{|N_T^2|}{|N_\mu^2|},
\end{equation}
for temperature gradient $\nabla \equiv d \ln P / d \ln T$ (with pressure $P$ and temperature $T$), adiabatic gradient $\nabla_\mathrm{ad}$, composition gradient $\nabla_\mu \equiv d \ln \mu / d \ln P$, $\phi$ and $\delta$ the logarithmic derivatives of density with respect to $\mu$ and $T$, respectively, and $N_T$ and $N_\mu$ the local Brunt–V\"{a}is\"{a}l\"{a} frequencies based on $T$ stratification alone and on $\mu$ stratification alone, respectively. 
A given layer is unstable to convection for $R_0 < 1$, and (in periodic or sufficiently tall domains) $1 < R_0 < \tau^{-1}$ becomes the necessary condition for instability to salt fingers, i.e., thermohaline convection \citep{baines_gill_1969,garaud_DDC_review}. 
All of our simulations presented here are run with $R_0 = 1.95$, corresponding to an initial $C$ profile that, for $0 \leq z \lesssim 100$, is far from marginal stability for both values of $\tau$ that we consider. 

\begin{figure*}
     \centering
     \includegraphics[width=1\linewidth]{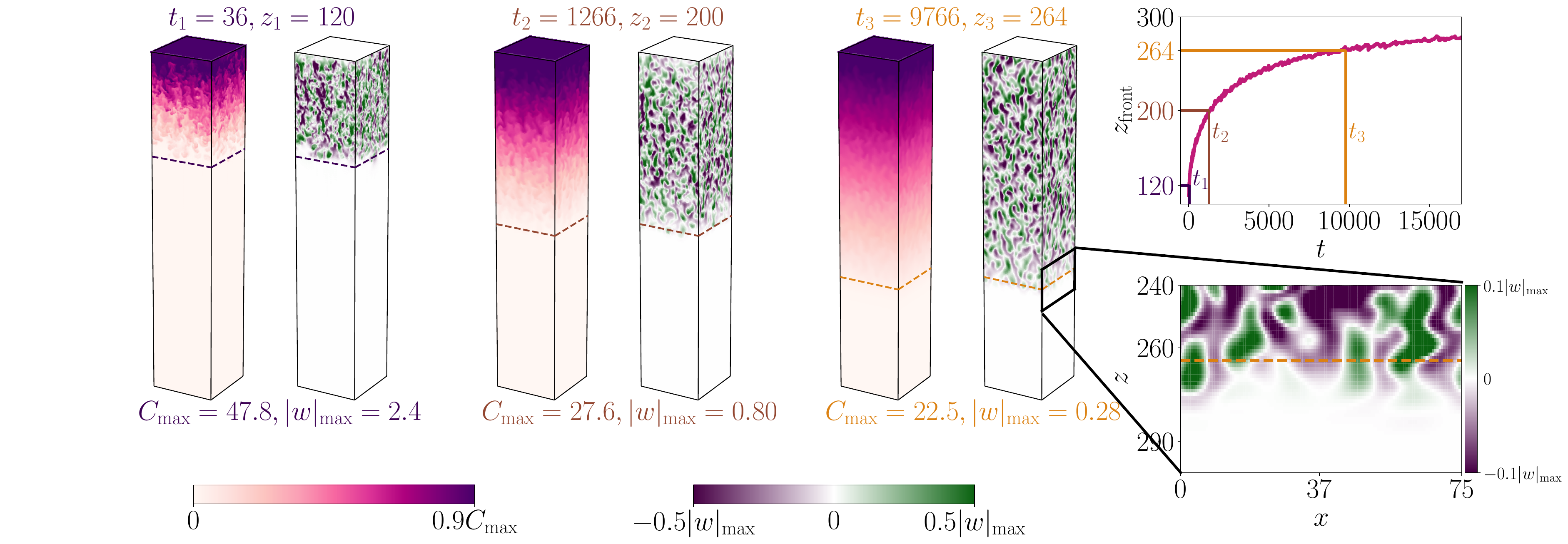}
     \caption{Composition (left) and vertical velocity (right) is visualized for Pr=0.05, $\tau=0.05$ at different times. The top right plot shows depth of the front (the boundary between thermohaline-stable and -unstable regions) as a function of time, with this depth also indicated by the dashed line on the 3D renderings. The bottom-right plot shows a 2D slice of the vertical velocity, zoomed in to the front location to highlight that dynamics extend past the front in an overshoot region. }
     \label{dyn}
\end{figure*}

As the simulations progress and the horizontally averaged $T$ and $C$ profiles change, we characterize the local stability at each $z$ by defining the \emph{local} density ratio $R_\rho(z)$. In our choice of units, this becomes the ratio of the horizontally averaged temperature gradient to the horizontally averaged composition gradient (cf.~Eq.~[27] of \citealt{tulekeyev_constraints_2024}). The unstable region can then be understood as the extent of the domain where $R_\rho(z) < \tau^{-1}$, and we define the boundary (or ``front") between the unstable and stable regions $\zfront$ as the depth $z$ where $R_\rho = \tau^{-1}$. 
Finally, it is often useful to introduce the \emph{reduced} density ratio
\begin{equation}
    r \equiv \frac{R_\rho - 1}{\tau^{-1} - 1}
\end{equation}
as a simpler measure of how unstable a region is, with the thermohaline-unstable region $1 < R_\rho < \tau^{-1}$ now mapping onto $0 < r < 1$, 
with $r < 0$ indicating instability to convection, and $r \to 1$ representing the limit of marginal stability to salt-fingers. 

\section{Results}
Figure \ref{dyn} summarizes a simulation with $\Pr = \tau = 0.05$. The three pairs of panels on the left show the composition (pink/purple) and vertical velocity (purple/green) at three different times. Early in the simulation, the upper region of the domain is strongly unstable, and vigorous fingers are launched that rapidly drag high-$C$ material down into the stable region and bring low-$C$ material up. This turbulent composition flux increases the vertical extent of the thermohaline-unstable region over time. This is indicated by the horizontal dashed lines in the left panels, denoting the local stability boundary $\zfront$ where $R_\rho(z) = \tau^{-1}$, and in the top-right panel, which shows $\zfront$ over time. 

Contrary to the one-time instability picture advocated by \citet{Koester2015}, we see that new fingers continue to form as the front progresses, feeding off the unstable $C$ gradient and gradually moving the front deeper.  
Taking the finger lifetime as approximately the e-folding time of the fastest-growing fingers, the early stage of rapid evolution lasts roughly $100$ finger lifetimes (with e-folding time calculated using the initial $R_0$), and our simulations evolve further for more than an additional $100$ finger lifetimes (based on the final value of $R_\rho$)---thus, after hundreds of generations of fingers, the layer remains unstable and gradually progresses deeper into the star.
%The top-right panel shows how $\zfront$ evolves with time. 

While the vigorous fingers at the earliest stages of the simulation drive a relatively fast front propagation speed, the system quickly settles into a regime where successive generations of fingers gradually push the front deeper on a timescale that is slow relative to a thermal diffusion time. In this regime, new fingers continually form, though each individual finger drives less flux than the earliest ones, as indicated by the smaller vertical velocities in the left panels.
Despite evolving for $O(10^4)$ thermal diffusion times, this simulation never reaches a state where the driving $C$ gradient has been stabilized by fast mixing processes and settles into a diffusive equilibrium---instead, consistent with 1D stellar evolution models that account for thermohaline convection \citep{bauer_polluted_2019,Wachlin_WDs}, we find that the system continues to evolve on a slow timescale due to continual thermohaline convection and successive generations of salt-fingers. We also note that fluid motions extend beyond the local stability boundary in an overshoot-like region, as highlighted in the zoomed-in panel in the bottom-right of Fig.~\ref{dyn}.

%\textcolor{red}{[Probably a decent place to comment on box height effects]}
These dynamics are observed while the front $\zfront$ is far from the bottom boundary, and we have verified that they are robust to changes in domain height $L_z$, thus suggesting that our key results hold in more extended domains. We find that as the front approaches the bottom boundary $\zfront \to L_z$, the flux in the overshoot region interacts with the boundary such that the bulk ($z < \zfront$) dynamics and composition profile become affected by the presence of the boundary. %(specifically it reduces bulk $r$ and causes a more unstable composition gradient). 
Thus, we restrict all reported results to times before the composition flux (turbulent or diffusive) becomes non-trivial at a depth of $\Delta z = 25$ away from the bottom boundary. %$F_{\mathrm{diff}}$ becomes non-negligible at $z=375$.

%\textcolor{red}{[Not clear to me that we need to include/discuss bottom-right panel. Especially since you've shown that at extremely small $\Pr$ and $\tau$, as Pascale predicted, mixing beyond the front is quite small. But leaving this note here for now in case we change our minds.]}

\begin{figure*}
     \centering
     \includegraphics[width=0.95\linewidth]{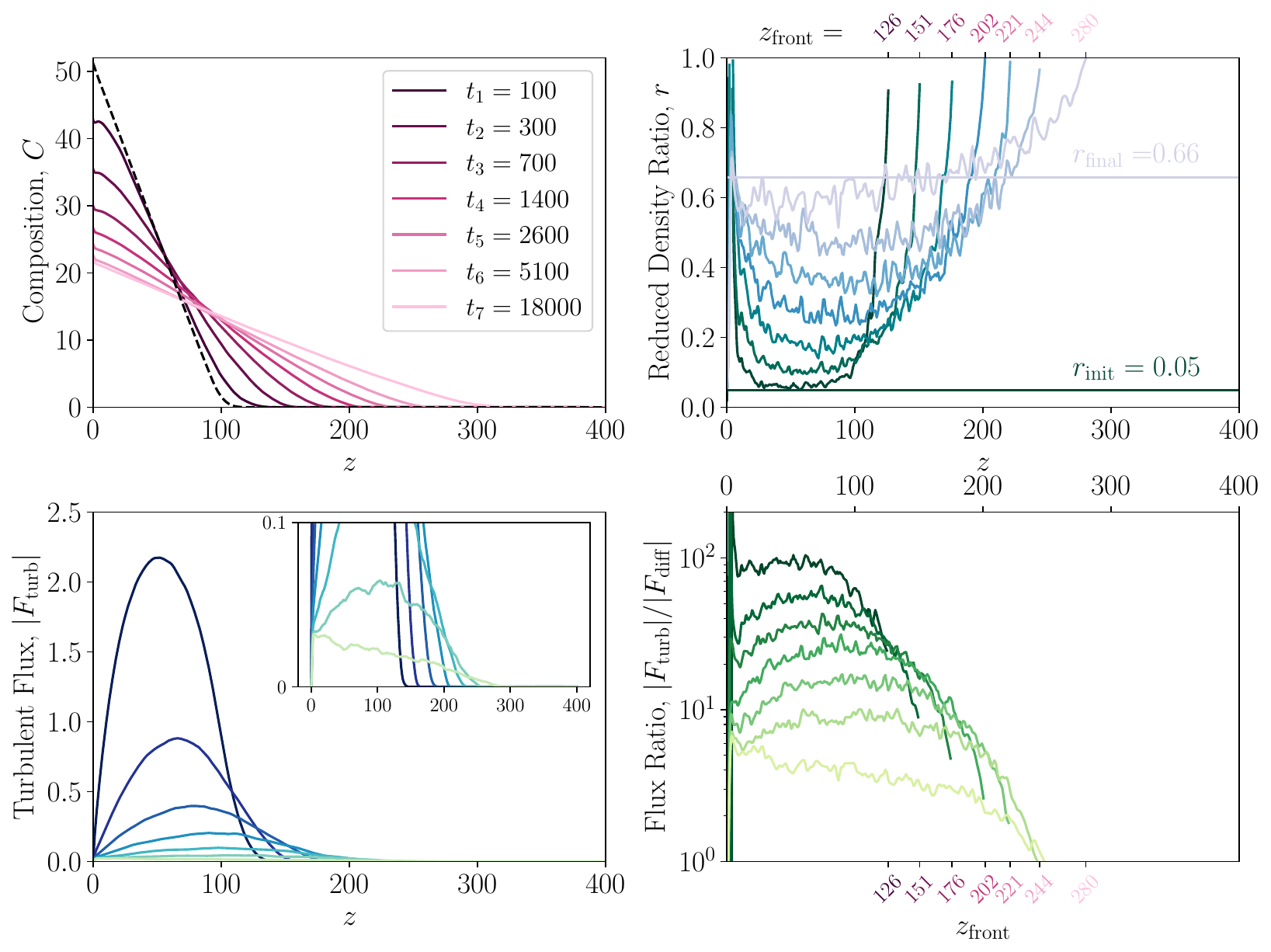}
     \caption{Vertical profiles of composition (top left), reduced density ratio (top right), advective flux (bottom left), and ratio of advective to diffusive flux (bottom right) at different times for the same simulation as in Fig.~\ref{dyn}. Lighter colors correspond to later times, and the dashed line in the top-left shows the initial profile. The thermohaline front propagates into the thermohaline-stable zone and mixes, partially relaxing the initial composition gradient. But due to the constant flux of high-metallicity material from the top boundary, the upper region remains unstable and drives more turbulent than diffusive flux. 
     }
     \label{prof}
 \end{figure*}

Additional details of the same simulation are shown in Fig.~\ref{prof}. Each panel shows horizontally averaged quantities at different times in the simulation, with lighter colors representing later times. Profiles are additionally smoothed by averaging in time over a window of width $\Delta t = 100$.  
The top-left panel shows the composition profile, which is seen to relax over time, as the extent extent of the region with non-zero composition gradient grows. From this horizontally averaged composition profile (and the corresponding temperature profile, not shown\footnote{Consistent with \citet{Brown_2013} (see their Appendix B) and \citet{xie_reduced_2017}, we find that the heat transport at low $\Pr$ and $\tau$ is dominated by diffusion, and thus the mean temperature profile remains largely identical to its initial linear diffusive equilibrium.}), we calculate the reduced density ratio $r$ from the local density ratio $R_\rho$ as described in Sec.~\ref{sec:methods} to measure the local stability of the profile at each depth $z$. The result is shown in the top-right panel of Fig.~\ref{prof} where, consistent with the propagation of the $C$ gradient to larger $z$, we see that the extent of the region with $r < 1$ (indicating instability to salt-fingers) grows with time. 

\begin{figure*}
     \centering
     \includegraphics[width=0.95\linewidth]{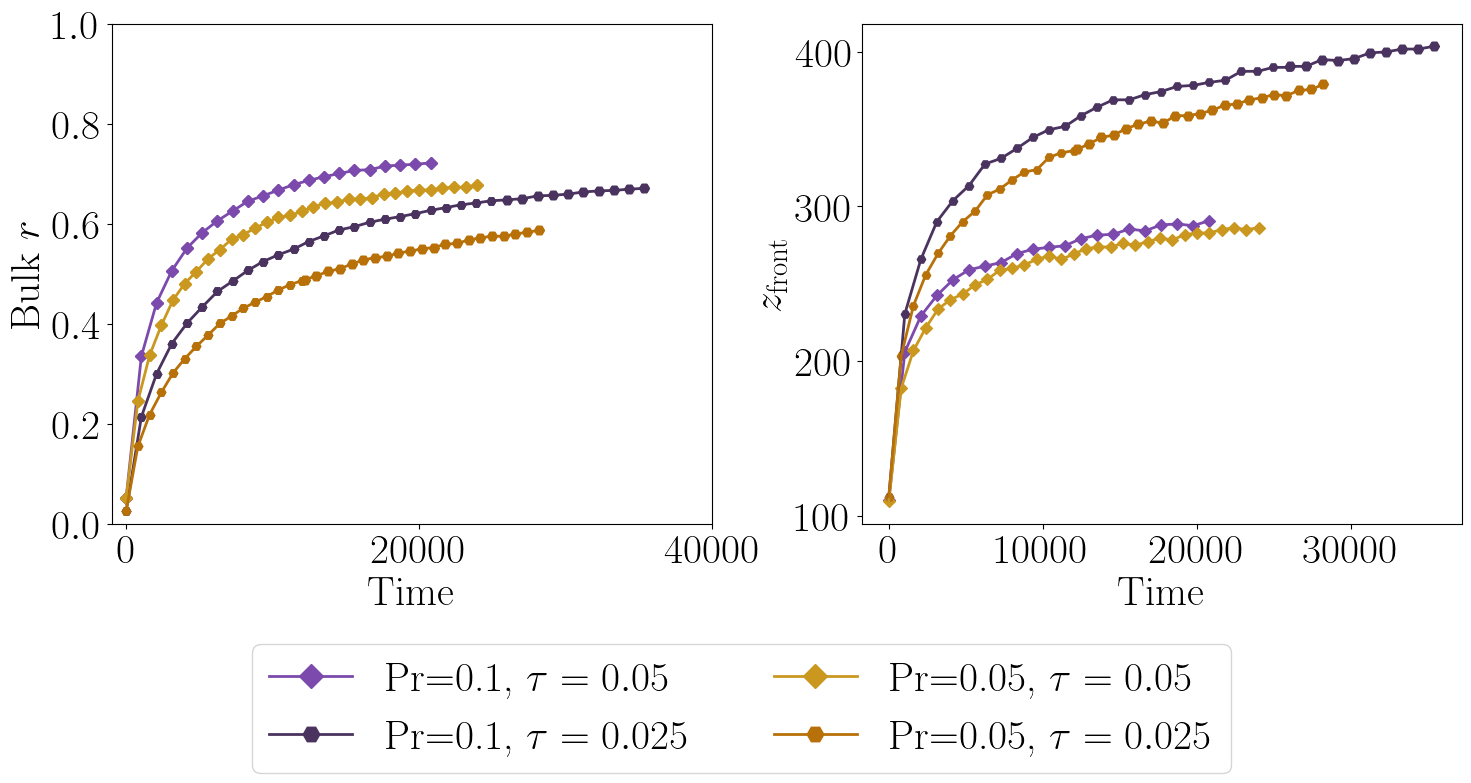}
     \caption{Bulk reduced density ratio $r$ (left) and front location $z_{\mathrm{front}}$ (right) over time for a set of simulations varying $\Pr$ and $\tau$. %Bulk values are calculated between $z=40$ and $z=z_{\mathrm{front}}-20$. 
     %The front location is where $R_{\rho}>1/\tau$. 
     %Reduced density ratio increases rapidly initially before tending towards a larger---but still unstable---value. 
     After a brief period of rapid increase, $r$ grows only very slowly and, importantly, remains below 1.
     Smaller Pr results in a smaller (more unstable) final $r$, with a similar trend in reducing $\tau$. Smaller Pr also results in faster front propagation. %Pr=0.05 runs are run at a larger box size of $Lz=500$ to allow a longer time for the front to propagate before it hits the bottom boundary. 
     }
     \label{bulkcomp}
 \end{figure*}

While this region does become less unstable over time, as shown by $r$ getting somewhat closer to 1, it never reaches marginal stability ($r = 1$). This is consistent with the dynamics shown in Fig.~\ref{dyn}, where fingers become less vigorous with time but do continue forming. This trend can also be inferred from the bottom-left panel of Fig.~\ref{prof}, which shows the vertical turbulent composition flux ($F_\mathrm{turb} = -\langle w C \rangle$ in these units, where $\langle . \rangle$ denotes a horizontal average) at different times. As the front progresses, significant turbulent flux occurs over a progressively larger region and, 
%the region over which significant turbulent composition flux occurs correspondingly grows and, 
consistent with the increase in $r$, the peak in the flux profile shrinks. 
%\textcolor{red}{[This sentence needs to change once we see the final fig 3]} 
However, as seen in the bottom-right panel showing the ratio of turbulent to diffusive composition fluxes, the turbulent flux always remains at least a factor of a few larger than the diffusive flux in the bulk (with the ratio getting smaller as $z \to \zfront$ where $r \to 1$, consistent with the expectation that $F_\mathrm{turb} \to 0$ as $r \to 1$). This suggests that the late-time configuration cannot be well-approximated by a profile governed by molecular diffusion alone and that models of polluted white dwarfs should account for thermohaline mixing when appropriate.

To check that these findings are not simply consequences of our choice of $\Pr$ and $\tau$, we show in Fig.~\ref{bulkcomp} how the bulk reduced density ratio $r$ (left)---defined as the average of $r(z)$ from $z = 30$ to $z = z_{\mathrm{front}}-20$ and front location $z_{\mathrm{front}}$ (right) vary with time across different values of $\Pr$ and $\tau$. We compare $\Pr = 0.1$ (purple) and $\Pr = 0.05$ (orange), as well as $\tau = 0.05$ (diamond markers) and $\tau=0.025$ (hexagons; note these runs use $L_z = 500$ to allow dynamics to evolve further before being significantly affected by the bottom boundary). Each run has an initial density ratio of $R_0=1.95$, corresponding to $r=0.05$ %for $\tau=0.05$ and $r=0.02436$ for $\tau=0.025$
and $r \simeq 0.024$ for $\tau = 0.05$ and $\tau = 0.025$, respectively. As the fronts propagate, each layer becomes less unstable. Initial increases in $r$ occur quickly and then settle on an unstable value as the fronts continue to propagate through the domain. 
%\todo{Should we say this is because of our simulation set up but we have reason to belive that's how it would happen irl?} [I don't think so]
Interestingly, runs with smaller $\Pr$ and $\tau$ generally result in smaller bulk $r$. That this occurs before any significant influence sets in from the (unphysical) bottom boundary suggests that this can be expected to occur in actual WDs. %While these simulations have not reached a steady state, one can imagine that if they were to run for a longer time with no physical bottom boundary and still have the input of unstable composition at the top boundary, that $r_f$ would settle on a final unstable value. Even if this value were to climb to stability, the timescale this happens is not instantaneous. 

%advective flux (bottom left), and vertical velocity $w_{\mathrm{rms}}$ (bottom right) vary with time for different parameters. We simulate Pr=0.1 (purple) and Pr=0.05 (orange/yellow) with diamond markers for $\tau=0.05$ and hexagon markers for $\tau=0.025$. In the top left plot of \ref{bulkcomp} we show the evolution of the bulk density ratio. Each run had an initial $R_\rho=1.95$, but this value increases as the front propagates and mixes. 

%This result is robust against box size increases \textcolor{red}{[something to decide later: could show this in an appendix]} meaning the boundary is not the reason for the plateau. 

The right panel of Fig.~\ref{bulkcomp} shows the progression of $\zfront$ (the depth where $r(z) = 1$) over time. Lower values of $\tau$ result in a quicker propagation, while lower Pr results in a slightly slower propagation. The lower $\tau$ and Pr runs have a lower final $r$, thus are more unstable, causing the fronts to propagate faster. %Again, these result are robust to box size and so one can imagine that in a star the front would continue to propagate into the interior. \todo{transistion to fig 5 here somehow? not sure what else to say about this-- I could report the zfront $~ t^x$ ?}

%The bottom left plot of Figure \ref{bulkcomp} shows how the bulk advective flux \todo{change to wdwarf config} decreases as the front propagates and mixes, and also plateaus. The final bulk flux is dependent on $\tau$ and not on Pr since Pr=0.1 $\tau=0.025$ (light purple) and Pr=0.05 $\tau=0.025$ (light orange) follow the same trend. These runs settle on the lowest advective flux, whilst Pr=0.1, $\tau$=0.05 (medium purple), and Pr=0.05 $\tau=0.05$ (medium orange) have a slightly higher advective flux. Despite having a smaller advective flux, the lower $\tau$ fronts propagate the fastest.

%A similar trend can be seen in the velocity (bottom right) plot. The bulk velocity rapidly slows but then settles on a nonzero value. The velocity of the smaller $\tau$ runs is smaller than larger $\tau$. We have confirmed this follows the Brown prediction well, but it is apparent that it does not capture the speed of propagation entirely. This is due to the stiffness of the boundary resulting in different bulk vs boundary values.

%\subsection{Figure 5: bulk vs boundary values}
\begin{figure*}
     \centering
     \includegraphics[width=0.95\linewidth]{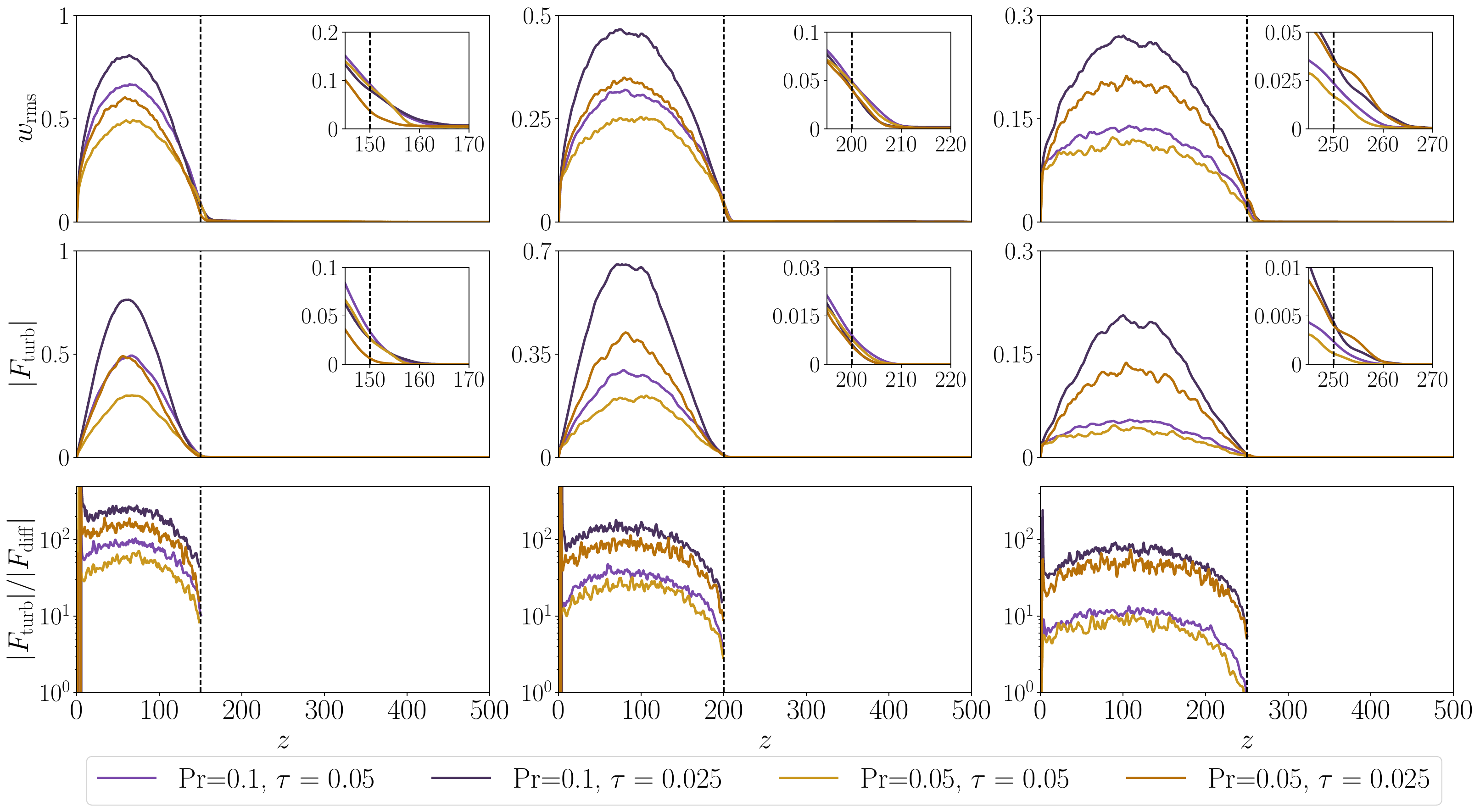}
     \caption{Horizontally averaged profiles of vertical velocity (top), turbulent composition flux (middle), and ratio of turbulent to diffusive flux (bottom) at different stages of front propagation for the same simulations as are shown in Fig.~\ref{bulkcomp}. 
     The black dashed line shows location of the front, $z_{\mathrm{front}}$, with insets focusing on this region. 
     Vertical velocity and turbulent fluxes show an overshoot region with fluid motions extending past the stability boundary and driving additional flux beyond the diffusive flux. 
     }
     \label{bvbcomp}
 \end{figure*}

Figure \ref{bvbcomp} shows horizontally averaged profiles of the rms vertical velocity $w_\mathrm{rms}$ (top), turbulent composition flux $|F_{\mathrm{turb}}|$ (middle), and ratio of turbulent to diffusive flux $|F_{\mathrm{turb}}|/|F_{\mathrm{diff}}|$ (bottom) for %$\Pr=0.1$ (purple) and $\Pr=0.05$ (orange)
the same simulations as in Fig.~\ref{bulkcomp}. Rather than comparing different simulations at the same time $t$, each column shows profiles with the same front locations $\zfront$, indicated by the black dashed line---and to help enable this comparison, the short time-averaging window is reduced from $\Delta t = 100$ (as in Fig.~\ref{prof}) to $\Delta t = 20$. From left to right, the columns correspond to $z_{\mathrm{front}}=150$, $z_{\mathrm{front}}=200$, and $z_{\mathrm{front}}=250$. Each inset zooms in on a narrow region around $\zfront$. 
The case corresponding to the fastest propagation ($\Pr = 0.1$, $\tau = 0.025$) is seen to have the largest bulk $w_{\mathrm{rms}}$ and $|F_\mathrm{turb}|$. 
Smaller $\Pr$ coincides with slightly lower bulk vertical velocities and turbulent fluxes and, correspondingly, slower propagation rates. % (though it does settle on a more unstable $r$ than the $\Pr = 0.1$ case). 
Note that these bulk values agree well with predictions based on $\Pr$, $\tau$, and $R_\rho(z)$ using the parasitic saturation theory described in \citet{Brown_2013}---further details provided in Appendix \ref{sec:Brown_comparison}. 
% However, while such parasitic saturation theories \citep[see also][]{RadkoSmith2012,Harrington_2019,fraser_magnetized_2024} accurately predict the turbulent fluxes in regions with $0 < r < 1$, these theories do \emph{not} predict any turbulent fluxes in regions that are locally stable. Indeed, while local simulations with imposed background gradients with $r > 1$ presumably exhibit no significant turbulent fluxes and thus are consistent with parasitic saturation models, the insets in Fig.~\ref{bvbcomp} demonstrate a non-local effect that is unaccounted for in such calculations: that fingers driven from unstable gradients near stable layers (i.e., for $z$ slightly less than $\zfront$, here) do not immediately stop upon reaching the stable layer, and instead are carried forward some distance by inertia and drive non-zero turbulent flux within the stable layer, much like convective overshoot \cite{Anders_review}. 
% \textcolor{red}{[Probably irrelevant but could point out that the Kippenhahn model does include some of this mixing]} While this effect appears to become less significant as $\Pr$ shrinks, and thus is likely insignificant in astrophysical parameter regimes, fingers often travel at approximately the Alfv\'en speed in the magnetized case \citep{Harrington_2019, fraser_magnetized_2024}, and thus might be expected to carry much more inertia as they hit $\zfront$ than their hydrodynamic counterparts.

\section{Conclusion}
Motivated by debate in the context of polluted white dwarfs as to the occurrence and role of thermohaline mixing (also called thermohaline convection or fingering convection), we have investigated the dynamics and longevity of salt fingers in a region with a destabilizing gradient of chemical composition $C$ (analogous to a $\mu$ inversion in a WD photosphere) overlying a stable layer, with a fixed flux of high-$C$ material (analogously, accreted heavy metals) from above. 

Contrary to the picture put forth in \citet{Koester2015} of thermohaline mixing as a one-time instability process that homogenizes driving gradients over some finite local thermal timescale, we demonstrate that thermohaline mixing persists over very long times (relative to a thermal diffusion time) when provided a constant flux of material, and that the depth of the thermohaline-unstable layer grows over time in a manner reminiscent of 1D models, such as those reported by \citet{Wachlin_WDs} (see their figures 3 and 4). While the driving $C$ gradient in this unstable layer does somewhat relax with time, it remains steep enough to continually drive new generations of unstable fingers. Importantly, the fluid does not appear to approach marginal stability ($r \to 1$) even over very long time scales, but instead settles into a configuration solidly in the regime of thermohaline instability, with a mixing front continuing to propagate deeper and deeper. Furthermore, even at these late times with relaxed gradients, the turbulent composition flux is always larger than the diffusive flux and is well-described by the \citet{Brown_2013} model, suggesting that late-time profiles are not merely determined by the diffusive equilibrium---or, if they do eventually reach a diffusive equilibrium, it takes an extreme amount of time relative to the relevant thermal diffusion time, raising the question of whether the diffusive equilibrium assumed in many polluted WD models is truly reached within the timescale of a given accretion event.

Our results broadly support the conclusions of \cite{Deal_WD,Wachlin2017,bauer_increases_2018,bauer_polluted_2019,Wachlin_WDs} and \cite{Dwomoh2023}, who have demonstrated that thermohaline mixing must be properly accounted for in hot WDs to accurately estimate mass accretion rates. Our simulations are limited by numerical constraints, most notably the need to use larger values of $\Pr$ and $\tau$ than are found in these stars and the constraint on the depth to which we can evolve our thermohaline regions before the artificial bottom boundary affects our dynamics. However, not only are we in the astrophysically relevant regime of $\Pr, \tau \ll 1$, but we have also varied $\Pr$ and $\tau$ and demonstrated that our results are robust to changes in these quantities and, most importantly, trend towards increased significance of thermohaline mixing as $\Pr, \tau \to 0$. Furthermore, our domain height $L_z$ is sufficiently tall to permit the system to evolve into one where the bulk properties evolve only very slowly if at all, lending credibility that our results would broadly hold for the case of $L_z \to \infty$.

We feel there is little ambiguity remaining as to whether thermohaline mixing is an important effect that must be included to accurately infer accretion rates in polluted WDs, unless some qualitatively new pieces of additional physics are introduced to the overall modeling picture that somehow quench this otherwise-robust instability. 
%Unless some qualitatively new pieces of additional physics are introduced to the overall modeling picture that somehow quench this otherwise-robust instability, thermohaline convection is
However, there are some avenues for future work remaining. Most notably, recent work has shown that magnetic fields can dramatically enhance the rate of thermohaline mixing in these systems \citep{Harrington_2019,fraser_magnetized_2024}, and can drive fingers to travel at roughly the Alfv\'en speed. This suggests, first, that the settling rate of pollutants likely depends strongly on the local magnetic field strength in WDs, and second, that the inertia and chemical flux carried by overshooting fingers (see bottom-right panel of Fig.~\ref{dyn}) may be much larger in the magnetized case, possibly significantly enhancing the rate at which the front propagates and presenting a non-local source of chemical mixing to possibly include in 1D models. This would imply even larger inferred accretion rates for polluted WDs, which may be problematic in terms of the overall picture of understanding how to supply such substantial amounts of material to WDs through asteriod accretion. 

\section*{Acknowledgements}
IGC, AEF, and BPB were partially supported by NASA HTMS grant 80NSSC20K1280. AEF was additionally supported by an NSF Astronomy and Astrophysics Postdoctoral Fellowship under award AST-2402142, the George Ellery Hale Postdoctoral Fellowship in Solar, Stellar and Space Physics at the University of Colorado, Boulder, and NSF award OCE-2023499. 
EHA was supported by NSF grant PHY-2309135 and Simons Foundation grant (216179, LB) to the Kavli Institute for Theoretical Physics (KITP).
BPB was partially supported by NASA grant HSR 80NSSC24K0270.

%[Make sure to acknowledge NASA Pleiades appropriately]
Computations were conducted with support of the NASA High End Computing (HEC) Program through the NASA Advanced Supercomputing (NAS) Division at Ames Research Center on Pleiades.

%[Include the KITP acknowledgement for transtar21, since that's where the idea for this originated.]
The authors acknowledge useful discussions with Pascale Garaud, Adam Jermyn, Matteo Cantiello, Rich Townsend, and Tim Cunningham, as well as the thermohaline working group at the KITP program “Probes of transport in stars” for fruitful discussion that led to the conception of this idea. 
This research was supported in part by the National Science Foundation under grant No.~NSF PHY-1748958.  % the line that KITP requests be included

\software{Dedalus v2 \citep{Dedalus_methods}, plotpal \citep{plotpal}, matplotlib \citep{Matplotlib}, NumPy \citep{numpy}, SciPy \citep{2020SciPy-NMeth}}

\clearpage

%% maybe a nice clearpage here for formatting depending on where length ends up?
%\clearpage
\appendix

\section{Governing equations, choice of units, and numerical methods} \label{sec:non-dimensionalization}
We simulate the governing equations for a stably-stratified fluid confined between two horizontal plates. We assume that flows are sufficiently slow compared to the local sound speed and that the vertical length scale is much smaller than a pressure scale height and the local radius of curvature of the star, under which the Boussinesq approximation is valid \citep{spiegel_boussinesq_1960} and we may take a local Cartesian coordinate system with $z$ increasing with depth. We neglect rotation and magnetic fields, and consider the effects on buoyancy of fluctuations in both temperature $T$ and chemical composition $C$ (analogous to mean molecular weight $\mu$).
%---where fluctuations are relative to a background linear profile $T_0$ and $C_0$---leading to the following governing equations in dimensional form \citep{garaud_DDC_review}:
While our setup (see Fig.~\ref{params}) differs slightly from the more common ``local" setup where double-diffusive convection is studied in a triply-periodic box with gradients in temperature and composition that are imposed and \emph{linear} \citep[see, e.g.,][]{RadkoSmith2012,garaud_DDC_review}, it is instructive to review these more standard setups before commenting on the minor differences to arrive at our setup, particularly because our choice of units is essentially identical. 

\subsection{Equations and units in the local case}
In these other setups with imposed background gradients, the governing equations in dimensional form are 
\begin{equation}
    \rho_m \left( \frac{\partial \mathbf{u}}{\partial t} + \mathbf{u} \cdot \nabla \mathbf{u} \right) = - \nabla p + \rho_m \nu \nabla^2 \mathbf{u} + \rho_m \left(-\alpha T + \beta C\right) \mathbf{g},
\end{equation}
\begin{equation}
    \frac{\partial T}{\partial t} + \mathbf{u} \cdot \nabla T + w \left( \frac{dT_0}{dz} - \frac{dT_\mathrm{ad}}{dz} \right) = \kappa_T \nabla^2 T,
\end{equation}
\begin{equation}
    \frac{\partial C}{\partial t} + \mathbf{u} \cdot \nabla C + w \frac{dC_0}{dz} = \kappa_C \nabla^2 C,
\end{equation}
\begin{equation}
    \nabla \cdot \mathbf{u} = 0,
\end{equation}
where $T$ and $C$ are departures about some background profiles. 
Here $dT_\mathrm{ad}/dz \equiv -g/c_p$ captures the effect of adiabatic cooling of a rising fluid parcel with $g = |\mathbf{g}|$ and $c_p$ the specific heat at constant pressure, $\rho_m$ is the mean mass density in the fluid layer, $\nu$ is the kinematic viscosity, $\alpha$ and $\beta$ are the coefficients of thermal expansion and compositional contraction, respectively, $\kappa_T$ is the thermal diffusion coefficient, and $\kappa_C$ is the compositional diffusion coefficient. Consistent with the Boussinesq approximation, $\nu$, $\kappa_T$, $\kappa_C$, $\alpha$, and $\beta$ are all assumed to be uniform and constant. The imposed background gradients of temperature and composition are given by $T_0$ and $C_0$ and are traditionally assumed to be strictly linear in $z$.

We follow \citet{garaud_DDC_review} in our choice of units. As our unit of length $[x]$, we take the typical finger width $d$, given by 
\begin{equation} \label{eq:x-units}
    [x] = d \equiv \left( \frac{\kappa_T \nu}{\alpha g \left( \frac{dT_0}{dz} - \frac{dT_\mathrm{ad}}{dz} \right)} \right)^{1/4} = \left( \frac{\kappa_T \nu}{N_T^2} \right)^{1/4},
\end{equation}
where $N_T$ is the local Brunt–V\"{a}is\"{a}l\"{a} frequency defined from the thermal stratification, and rather than $T_0$ representing an imposed background temperature gradient, we use the initial temperature gradient (see Fig.~\ref{params}). 
As our unit of time we take the thermal diffusion time across a width $d$, 
\begin{equation} \label{eq:t-units}
    [t] = \frac{d^2}{\kappa_T},
\end{equation}
and our units for temperature and composition are, respectively, 
\begin{equation}
    [T] = d \left( \frac{dT_0}{dz} - \frac{dT_\mathrm{ad}}{dz} \right), 
\end{equation}
and
\begin{equation}
    [C] = \frac{\alpha}{\beta} [T].
\end{equation}
%While our background temperature gradient $T_0$ is not fixed in our setup, we still take the initial temperature gradient (see Fig.~\ref{params}) in our choice of $[x]$ and $[T]$.

The non-dimensional governing equations are then
\begin{equation} \label{eq:dimless-mom-standard}
    \frac{\partial \hat{\mathbf{u}}}{\partial \hat{t}} + \hat{\mathbf{u}} \cdot \hat{\nabla} \hat{\mathbf{u}} = - \hat{\nabla} \hat{p} + \Pr \hat{\nabla}^2 \hat{\mathbf{u}} + \Pr \left( \hat{T} - \hat{C} \right) \hat{\mathbf{e}}_z,
\end{equation}
\begin{equation}
    \frac{\partial \hat{T}}{\partial \hat{t}} + \hat{\mathbf{u}} \cdot \hat{\nabla} \hat{T} + \hat{w} = \hat{\nabla}^2 \hat{T},
\end{equation}
\begin{equation} \label{eq:dimless-C-standard}
    \frac{\partial \hat{C}}{\partial \hat{t}} + \hat{\mathbf{u}} \cdot \hat{\nabla} \hat{C} + \frac{\hat{w}}{R_0} = \tau \hat{\nabla}^2 \hat{C},
\end{equation}
and
\begin{equation}
    \hat{\nabla} \cdot \hat{\mathbf{u}} = 0.
\end{equation}
Here, $\hat{\mathbf{e}}_z$ is the unit vector in the $z$ direction, and hats over fields, coordinates, and derivatives indicate non-dimensional quantities---note hats are suppressed in the main text. %\textcolor{red}{[phrasing here will depend on whether axis labels are things like ``time" or $t/[t]$]}. 

The above system is specified by three dimensionless control parameters: 
\begin{equation} \label{eq:control-parameters}
    \Pr = \frac{\nu}{\kappa_T}, \quad \tau = \frac{\kappa_C}{\kappa_T}, \quad \text{and} \quad R_0 = \frac{\alpha \left( \frac{dT_0}{dz} - \frac{dT_\mathrm{ad}}{dz} \right)}{\beta \frac{dC_0}{dz}},
\end{equation}
where $\Pr$ is the Prandtl number and $\tau$ is the diffusivity ratio, both of which are extremely small in stars \citep[see, e.g.,][]{Garaud2015}. In local models, the density ratio $R_0$ characterizes the stabilizing influence of the background temperature gradient relative to the destabilizing composition gradient and, in radiation zones, can take any value from 1 to infinity depending on local stellar conditions. The threshold $R_0 = 1$ is equivalent to the Ledoux criterion for convection, and layers with $1 < R_0 < \tau^{-1}$ are stable to convection but generally unstable to salt-finger motions \citep{baines_gill_1969}. %, as described below in Sec.~\ref{setup:subsec:linear}.
In the case of imposed background gradients, $R_0$ appears in equation Eq.~\eqref{eq:dimless-C-standard} because it is precisely $dC_0/dz$ measured in the units $[x]$ and $[C]$---i.e., the term with $R_0$ appearing in Eq.~\eqref{eq:dimless-C-standard} could be written instead as $\hat{w} d\hat{C}_0/d\hat{z}$ with the understanding that, in these units, $d\hat{C}_0/d\hat{z} \equiv 1/R_0$. %In our setup, there is no imposed background $dC_0/dz$, so there is no explicit $R_0$ app

\subsection{The case without fixed background gradients}
Rather than impose fixed linear temperature and composition profiles, we merely begin with initial temperature and composition distributions as part of our initial conditions on $T$ and $C$, and allow them to evolve freely in response to the effects of advection and diffusion. Our non-dimensional governing equations are thus
\begin{equation} \label{eq:dimless-mom}
    \frac{\partial \hat{\mathbf{u}}}{\partial \hat{t}} + \hat{\mathbf{u}} \cdot \hat{\nabla} \hat{\mathbf{u}} = - \hat{\nabla} \hat{p} + \Pr \hat{\nabla}^2 \hat{\mathbf{u}} + \Pr \left( \hat{T} - \hat{C} \right) \hat{\mathbf{e}}_z,
\end{equation}
\begin{equation}
    \frac{\partial \hat{T}}{\partial \hat{t}} + \hat{\mathbf{u}} \cdot \hat{\nabla} \hat{T} = \hat{\nabla}^2 \hat{T},
\end{equation}
\begin{equation} \label{eq:dimless-C}
    \frac{\partial \hat{C}}{\partial \hat{t}} + \hat{\mathbf{u}} \cdot \hat{\nabla} \hat{C} = \tau \hat{\nabla}^2 \hat{C},
\end{equation}
and
\begin{equation} \label{eq:dimless-incompressible}
    \hat{\nabla} \cdot \hat{\mathbf{u}} = 0,
\end{equation}
where $\hat{T}$ and $\hat{C}$ are initialized with the profiles shown in Fig.~\ref{params}. %, and small-amplitude noise is added to $\hat{T}$ to seed instability. 
We choose the same units as in the local case, except now $dT_0/dz$ is interpreted as the initial temperature gradient rather than an imposed one. With this setup, $1/R_0$ no longer appears explicitly in our equations, and instead is the value of the initial $d\hat{C}_0/d\hat{z}$ in the top portion of our domain. Finally, we impose periodic boundary conditions in the horizontal, and in the vertical we impose no-slip boundary conditions, with $d\hat{T}/d\hat{z}$ and $d\hat{C}/d\hat{z}$ fixed at $\hat{z}=0$ and $\hat{T}$ and $\hat{C}$ fixed at $\hat{z}=L_z$.

\subsection{Numerical methods}
We solve equations \ref{eq:dimless-mom}-\ref{eq:dimless-incompressible} using the pseudospectral-tau method implemented in Dedalus v2 \citep{Dedalus_methods} with a 3rd-order semi-implicit Adams-Bashforth/BDF timestepping scheme \citep[see][Eq.~2.14]{Wang_timesteppers_2008} where our step size is restricted by an advective CFL with safety factor 0.2, and is restricted by the period of an internal gravity wave\footnote{Note that internal gravity waves are often very important in thermohaline convection \citep{Stellmach2011} and thus, even if the corresponding terms are timestepped implicitly, it is often crucial to ensure they are timestepped \emph{accurately}, not just stably. That said, \citet{xie_reduced_2017} demonstrate that in regimes where $\tau \ll 1$ and $R_0\tau = O(1)$ (thus, the late times in all simulations reported here), these same terms become asymptotically small and thus this restriction on the step size may be unnecessary.}. 
All fields are expanded as a Fourier series with $N_x = N_y = 128$  Fourier modes in the $x$ and $y$ directions, respectively, and $N_z = 512$ or $N_z = 1024$ Chebyshev polynomials in $z$ for runs with $L_z = 400$ and $L_z = 500$, respectively. To avoid aliasing errors, we use the 3/2 dealiasing rule so that nonlinearities are evaluated in real-space on a grid of $192$ points in the horizontal and $768$ (for $L_z = 400$) or $1536$ (for $L_z = 500$) points in the vertical. 
Convergence checks were performed on the most numerically-taxing simulations by doubling each resolution and verifying that the dynamics are qualitatively identical and that the bulk properties and their evolution were unchanged. 
Additionally, recognizing that horizontal domain sizes ($L_x$ and $L_y$) can significantly impact low-$\Pr$ fingering convection if $L_x$ or $L_y$ are too small \citep{garaud_2d_2015}, a subset of simulations were checked by doubling $L_x$ and $L_y$ from our fiducial values of $75$ to $150$ (while also doubling $N_x$ and $N_y$) to ensure dynamics are not sensitive to domain width effects. 
Our initial conditions are $\hat{\mathbf{u}} = 0$, and $\hat{T}$ and $\hat{C}$ shown in Fig.~\ref{params} (specifically, $\hat{T} = \hat{z} - L_z/2$, and $d\hat{C}/d\hat{z}$ given by a shifted error function that transitions from $R_0^{-1}$ to 0 over a characteristic width of $\Delta \hat{z} = 10$ centered about $\hat{z} = 100$), with the addition of a small-amplitude random noise perturbation to $T$ of magnitude $10^{-3}$ to seed instability. %The code used to perform all simulations can be found \textcolor{red}{here}.
\section{Comparison to BGS13 model} \label{sec:Brown_comparison}

\begin{figure*}
     \centering
     \includegraphics[width=0.95\linewidth]{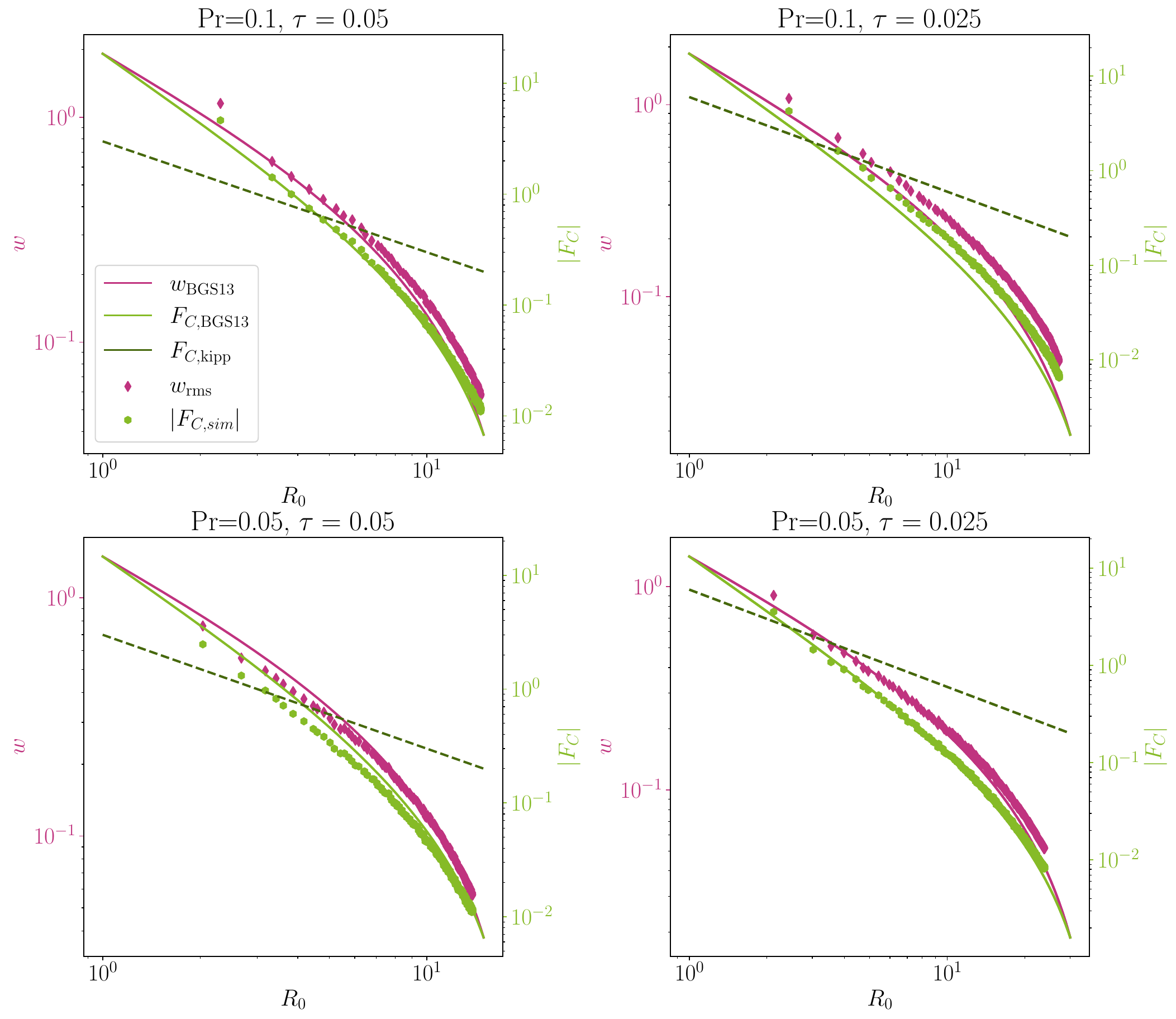}
     \caption{Vertical velocity (pink) and turbulent composition fluxes (green) vs $R_0$ for simulation data (points) compared to BGS13 (solid) and Kippenhahn (dashed) model predictions.}
     \label{brown}
\end{figure*}

Figure \ref{brown} compares simulation data (points) to the predictive model put forth in \citet{Brown_2013} (henceforth ``BGS13") (solid lines) for bulk rms vertical velocity $w_{rms}$ (pink) and bulk turbulent composition flux $|F_C|$ (green) as functions of bulk density ratio $R_\rho$, where ``bulk" quantities are averaged from $z=30$ to $\zfront - 20$. The flux predicted by the Kippenhahn model is shown by the dashed line where $\alpha_{\mathrm{th}}=0.5$. Clearly, the BGS13 model predicts bulk quantities extremely well, as expected from previous studies in triply-periodic domains \citep{Brown_2013,garaud_DDC_review,fraser_magnetized_2024}. For a more thorough comparison between the BGS13 model and a simulation where the local density ratio similarly varies with depth, see \citet{zemskova_fingering_2014}.
%Previous studies comparing 3D simulations to the BGS13 model are done in a triply periodic domain with no moving front \textcolor{red}{is this true, Pascale did a front at somepoint right?}.

\bibliography{thermohaline_refs}{}
\bibliographystyle{aasjournal}

\end{document}